\documentclass[conference]{IEEEtran}
\usepackage{blindtext, graphicx, color}
\usepackage{soul}
\usepackage{url}
\usepackage[normalem]{ulem}
\usepackage{subcaption}
\ifCLASSINFOpdf
\else
\fi
\begin{document}
%
\title{Holistic Approach for  Fault-Tolerant Network-on-Chip based Many-Core Systems}

\author{\IEEEauthorblockN{Siavoosh Payandeh Azad, Behrad Niazmand, Jaan Raik, Gert Jervan, Thomas Hollstein}
\IEEEauthorblockA{Department of Computer Engineering \\
Tallinn University of Technology\\
Email: \{siavoosh, bniazmand, jaan.raik, gert.jervan, thomas\}@ati.ttu.ee
}

}


%


\maketitle

\begin{abstract}
In this paper we describe a holistic approach for Fault-Tolerant Network-on-Chip (NoC) based many-core systems that incorporates a System Health Monitoring Unit (SHMU) which collects all the fault information from the system, classifies them and provides different solutions for different fault classes. A Mapper/Scheduler Unit (MSU) is used for online generation of different mapping and scheduling 
solutions based on the current fault configuration of the system. For detection of faults, we have leveraged concurrent online checkers, able to capture faults with low detection latency and providing the fault information for SHMU, which can be later used for the recovery process. The experimentation setup is performed in an open source tool, able to perform the mapping, scheduling and simulation of the system.

\end{abstract}


%
\IEEEpeerreviewmaketitle

\section{Introduction}

As the integration density of the cores on a single chip increases, it can cause many-core Network-on-Chip (NoC) based systems to be more susceptible to different sources of faults, such as Electro-Migration (EM), wear-out, etc, thus affecting their dependability. Even if all the manufacturing defects are captured during initial testing, faults that occur during the lifetime of the device are inevitable. Therefore, there is a necessity to address faults during the lifetime of the system. The growing size of Dark-Silicon on the chip, provides us with the advantage of having redundant units on the chip which can be exploited for mitigation of such faults. In this paper, we introduce an architecture for fault-tolerance and fault management in many-core NoC based systems. 

The proposed approach, equips the system with different components for detection, classification and recovering from faults. This is performed both at hardware level and system and application level. To this end, the system is equipped with a System Health Monitoring Unit (SHMU) which keeps a holistic view of the system's health status in a System Health Map (SHM) memory. SHMU collects the fault information via iJTAG \cite{IJTAG} mechanism from the system components. In order to detect faults, each router in the NoC is augmented with a fault detection mechanism implemented as concurrent online checkers \cite{CHECKERS}. This would provide a holistic view of the system's health status to SHMU in order to take further actions at system and application level.  

In order to take care of faults at the system and application level, SHMU utilises a Mapper/Scheduler Unit (MSU) in the system, which is in charge of mapping and scheduling application tasks onto the Processing Elements (PEs). SHMU communicates with MSU via a shared memory to pass the current state of the system to be used by mapping heuristics. For instance, if a permanent fault occurs on one of the PEs, it would make it possible to perform re-mapping and/or re-scheduling of the tasks, based on different heuristics implemented in MSU. In addition, in order to avoid packets with unreachable destinations being injected into the network and thus wandering in the system and causing conditions such as congestion or never reaching their destinations, the system uses NoCDepend \cite{NOCDEPEND1} mechanism which keeps compact information (in form of rectangle regions) about the non-reachable areas of the NoC in each router's output, providing routers with a holistic view of such non-reachable areas.  In this case if the destination is not accessible from any of the router's outputs, the packet would be dropped instantly, preventing all further hazards in the system.

The rest of this paper is organized as follows. In section \ref{RelatedWork}, the state-of-the-art literature is discussed. Section \ref{SystemArch} is dedicated to describing the overall system architecture and sections \ref{ArchUnderUse} to \ref{MSU} focus on each component separately. In Section VII, description of the experimentation setup using a developed open source tool is provided. Finally, Section \ref{Conclusion} concludes the paper, along with discussions about future plans.

\section{Related Work}
\label{RelatedWork}

In this section, the state-of-the-art research which address  faults during the lifetime of a NoC-based many-core system is reviewed. Different works have addressed faults and fault-tolerance in NoCs from a specific point of view, each one considering a different fault model. From one main perspective, the current works in the literature can be categorized as architecture-level and system-level approaches in dealing with faults. Indeed, other classification criteria are also addressed in different works, such as classifying faults by their occurrence frequency to transient, intermittent or permanent. In terms of level of granularity, different works have addressed faults on communication links, on-chip routers and turns and on-chip cores (Processing Elements) or a combination of them. 

In general, architecture-level approaches usually impose additional area overhead in order to tolerate faults, whereas system-level and application-level approaches tackle the issue by delegating the task more to the software side, such as re-mapping applications onto the network or applying redundancy schemes in software. Usually, system-level approaches incur negligible area overhead to the system, which makes them a better option. In the following, some of the architecture-level and system- and application-level approaches for handling faults in Network-on-Chips are reviewed. 

\subsection{Architecture-level approaches}

Architecture-level approaches  address faults by augmenting the network with additional components. For instance in \cite{FT-TOPOLOGY,ELASTINOC,FT-MESH-3D,FT-MESH,SPARE,SPARE2,SPARE3} fault-tolerant network topologies and redundant components (such as spare cores or routers) are introduced in order to achieve fault-tolerance. \cite{FL2STAR} has applied redundancy to links in order to tolerate link failures. Tosun et al. have introduced a fault-tolerant irregular topology generation method for application-specific Network-on-Chips \cite{FT-TOPOLOGY}, pursuing two main goals: (1) generating a fault-tolerant topology that would be able to maintain communication between tasks of an application despite of some link failures in the network and (2) proposing a Simulated Annealing (SA) based mapping algorithm for minimizing total energy consumption as the cost function. The scope of \cite{FT-TOPOLOGY} is focusing maily on link failures, therefore, faults in Processing Elements (PEs) and router turns are not addressed. Moreover, single link failure is considered, although it is mentioned that multiple links errors can also be tolerated as long as the failure of multiple links does not disconnect the network. 

In \cite{SPARE3}, a fault-tolerant reconfigurable NoC architecture is proposed which addresses fault tolerance by router redundancy. To be more specific, each Processing Element (PE) is connected to two routers. The proposed approach works based on the concept of hardware redundancy with spares, thus in each redundant router group, if one gets faulty, the spare one takes the role of the faulty one, thus guaranteeing that the network continues its normal operation. However, two drawbacks of this work would be that (1) considering spare router for each Processing Element (PE) in the network can impose significant area overhead, specially that the spare one is not used during the normal functioning of the network, and (2) in this work only fault in one router is considered, and occurrence of fault on both routers connected to a PE is not investigated.   

Some other works have tried to tackle faults in routers or links via fault-tolerant routing algorithms, for instance \cite{FT-IRREGULAR,FT-ROUTING-2D} for 2D Mesh based NoCs and  \cite{FT-ROUTING7,FT-ROUTING8,ELEVATOR-FIRST} for 3D NoCs derived from the 3D Mesh with limited number of vertical links in each layer. Bishnoi et al. have proposed a mechanism for implementing routing algorithms in a fault-tolerant manner for 2D Mesh NoCs and networks with topologies derived from the 2D Mesh due to faults in the network \cite{FT-IRREGULAR}. The proposed approach does not require routing tables at each router and the routing logic is constant as the network scales up. Moreover, it does not depend on the switching mechanism used in the router, thus it supports both wormhole and virtual cut-through switching mechanisms. One limitation of this work is that for analysis of the proposed mechanism, only one and two faulty link scenarios are considered.

In most of the works dealing with architecture-level approaches, there is usually a trade-off between the area overhead and the robustness of the approach in order to cover more faults. Some approaches rely on using Virtual Channels (VCs) in order to avoid the formation of deadlock, such as \cite{FT-ROUTING,FT-ROUTING9,ELEVATOR-FIRST}, whereas some others intend to save area consumption by avoiding the use of VCs (e.g. \cite{ROUTING_WITHOUT_VC}) , however, they instead restrict the routing turns and/or consider some simplifying assumptions regarding occurrence of faults (such as single faults) or existence of specific links (known as pillar) in case of 3D NoCs. 

Reconfigurability is also one of the trends in order to tolerate faults at router or link level, such as the work in \cite{ULBDR}. Despite the advantages, architecture-level methods impose additional area overhead and some of them are limited to specific fault models and are not generic, i.e. they do not address faults simultaneously on links, cores (PEs) and routing turns.

\subsection{System-level and Application-level approaches}

On the other hand, system-level and application-level approaches exist for handling faults. Such approaches mostly put the burden on software and try to provide flexibility while not letting the system performance degrade significantly when tolerating faults. System-level approaches also have different levels of granularity when addressing faults, either considering faults on PEs, routing turns and/or communication links. Some of them focus on providing fault-resiliency via re-mapping and/or re-scheduling tasks onto the network, such as \cite{FARM}. Most works have tried to perform the re-mapping offline, thus not imposing run-time latency overhead. Some works also use redundancy techniques, but at the software level in order to tackle faults. In total, system-level and application-level approaches are more efficient compared to architecture-level ones in terms of area overhead and run-time performance degradation. In the following, some of worked regarding system-level and application-level approaches are reviewed. 

In \cite{SHIFA}, authors have introduced SHiFA, which is a system-level approach for tolerating faults at run-time without imposing hardware overhead. The work addresses faults on links, Processing Elements (PEs) and routers. The proposed approach guarantees that mapping of applications is performed on healthy nodes. The design of SHiFA is based on distributed operating systems, thus being scalable for large networks. SHiFA uses a hierarchical kernel structure which based on the kernel's duty, it can be either a Mobile Master (MM) (keeps a view of the system and administers the Application Managers), Application Manager (AM) (capabale of mapping tasks onto nodes with basic kernel), or basic kernel. Despite the distributed characteristic and scalability advantages SHiFA possesses, if the Mobile Master fails (MM) further care should be taken in order to delegate the task to another node, however, this is left as future work.

Authors of \cite{FARM} have proposed a system-level Fault-Aware Resource Management (FARM) approach, which tends to tackle permanent, transient and intermittent faults via fault-aware mapping algorithm. The ultimate goal of the mapping scheme is to optimize system performance and energy consumption. One of the limitations of this work is that its focus is mainly on occurrence of faults on Processing Elements (PEs). To this end, spare cores (PEs) are utilized in order to delegate the duty of the failed cores to them when re-mapping. Thus, the problem of mapping is translated into the problem of spare cores placements (as they might not be needed at every router in the NoC) and assignment of tasks to them. One disadvantage of this work could be that leveraging spare cores imposes additional area overhead to the system, while the faults could have been handled mostly at application or system-level without additional significant hardware required. 

In \cite{FT-NORCHIP}, a fault-tolerant approach for application-specific NoCs is introduced. The methodology works based on constructing a decision tree for the different paths in the network, which is used to determine alternative paths in case one becomes faulty. The goal is to both improve reliability and traffic distribution in the network. To this end, the approach is described using problem formulation, while describing the network using graph theory, and it is also claimed that the approach can be integrated with any topology and any routing algorithm. Although the usage of routing tables allows implementation of any routing algorithm in the network, it suffers from scalability issues. As stated in the paper, the proposed method covers faults only on communication links, but no information is provided regarding addressing faults on Processing Elements (cores) and routing turns. 

Manolache et al. have proposed a heuristic-based approach \cite{FT-ENERGY-MAPPING}, which is based on message redundancy in order to address faults on communication links and also aim to minimize communication energy while meeting the stringent time-constrained characteristics of applications. The proposed approach tries to provide a guaranteed message arrival probability and guaranteed worst case response time. Despite the advantages, the proposed approach imposes information overhead as multiple copies of messages need to be created in order to tolerate faulty network links (achieved via temporal and spatial redundancy). Moreover, faults on routing turns and Processing Elements (PEs) are not addressed in this work. 

The work in \cite{FAULT-PREDICT} has presented the addition of a Fault Prediction Module (FPM) to the Phoenix NoC architecture, which makes it possible to predict the occurrence of faults based on their tendency and it is controlled by a threshold value. The focus is on fault prediction on network links, which can be either operating properly, operating with fault tendency or with permanent fault. It is shown that the area and power  overhead of the proposed module is acceptable compared to the other compared works. It is worth noting that the detection of faults is performed by a fault monitor unit and the classification of the fault tendency is done via the software part located in the Processing Element (PE), named OSPheonix. One limitation of this work would be that for the experiments, only single and double fault scenarios are considered and thus more number of faults and also faults on Processing Elements (PEs) and routing turns are not addressed. In addition, considering three counters per each input port of each router in the network (except the local port) imposes additional area overhead to the system.

The proposed approach in our work incorporates both system- and application-level and architecture-level approaches in order to provide a hybrid solution, thus making it possible to perform run-time fault-aware management of NoC-based many-core systems.

\section{System Architecture}
\label{SystemArch}

\begin{figure}
\centering
\includegraphics[width=\columnwidth]{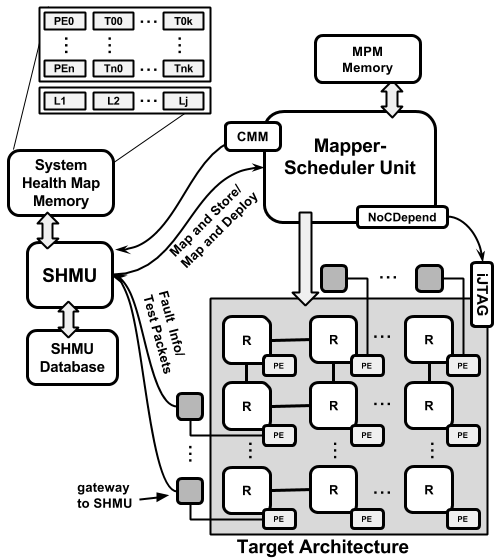}
\caption{Proposed approach for adding fault tolerance mechanisms}
\label{fig:SystemArch}
\end{figure}

In this paper, we propose an architecture for a fault-tolerant many-core Network-on-Chip based system. The important components of the architecture are shown in Fig. \ref{fig:SystemArch}. In this section we describe these components briefly.

\subsection{Target Architecture} 
The target architecture is a many-core NoC based system that has one Router and one Processing Element (PE) in each NoC tile. This can be either a 2D or a 3D NoC topology. Our assumption is that the architecture is prone to faults and would be either heterogeneous\footnote{By the term heterogeneous throughout this paper we mean the situation in which faults degrade the network and disable links, Processing Elements, turns in routers or affect some components of the network.} from the beginning by design or due to manufacturing defects. During its lifetime, the architecture will be affected by wear out and permanent faults. 

\subsection{System Health Monitoring Unit (SHMU)} 

SHMU is responsible for tracking diagnosis information in the system, classification and abstraction of this information, evaluation of the impact of the faults and updating System Health Map (SHM). The fault information collected from the system by using a fault detection mechanism provides SHMU with a holistic view of system components' health status. In order to handle faults, SHMU interacts with other modules in the system as well which are described as follows.  

\subsection{System Health Map (SHM)}
 System Health Map (SHM) is a shared memory between SHMU and Mapper-Scheduler Unit (MSU), which keeps the fault information about each component. 
This fault information is very abstract and includes all the turns in the router, nodes' health status (faulty or non-faulty) and nodes' speed degradation due to aging. This means that for each network tile $i$, SHM keeps the $PE_i$ health value along with the abstract health of the router's control part by assigning a health value to turns $T_{i,k}$, where $k$ denotes different 90 degree turns in the same router. On top of these, each physical link between routers has a health value $L_j$ assigned to it in the System Health Map memory. These values are binary and only show if a unit is faulty or non-faulty. Finer granularity of health representation can be taken into account, but it adds significantly to the size of the SHM.  SHM also keeps PE's frequency decrement due to aging in a byte. The size of SHM would be dependent on the size and topology of the target architecture.

SHMU writes into SHM and can read from it, however, the MSU can just read the contents of SHM. System Health Map can be manipulated by SHMU in order to generate some specific mapping based on predictions made. Fig. \ref{fig:SHMandRG}a shows visualization of an example of SHM for a $2\times2$ 2D Mesh network with XY routing algorithm. Each router has eight turns that are either red (restricted) or black (allowed) turns. The edges between two routers are representing the physical links and the small circle at the southern east of each router, represents the Processing Element (PE) in that tile. 

\subsection{Mapper-Scheduler Unit (MSU)} 

\begin{figure}
\centering
\includegraphics[width=\columnwidth]{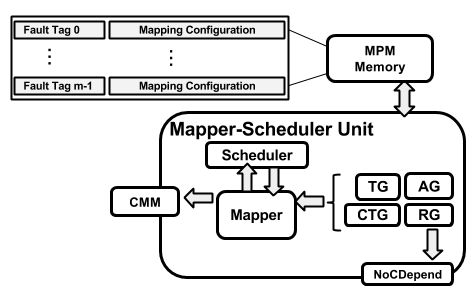}
\caption{Architecture of Mapper/Scheduler Unit}
\label{fig:MSU}
\end{figure}

\begin{figure}
\begin{subfigure}{.49\columnwidth}
  \centering
  \includegraphics[width=0.95\columnwidth]{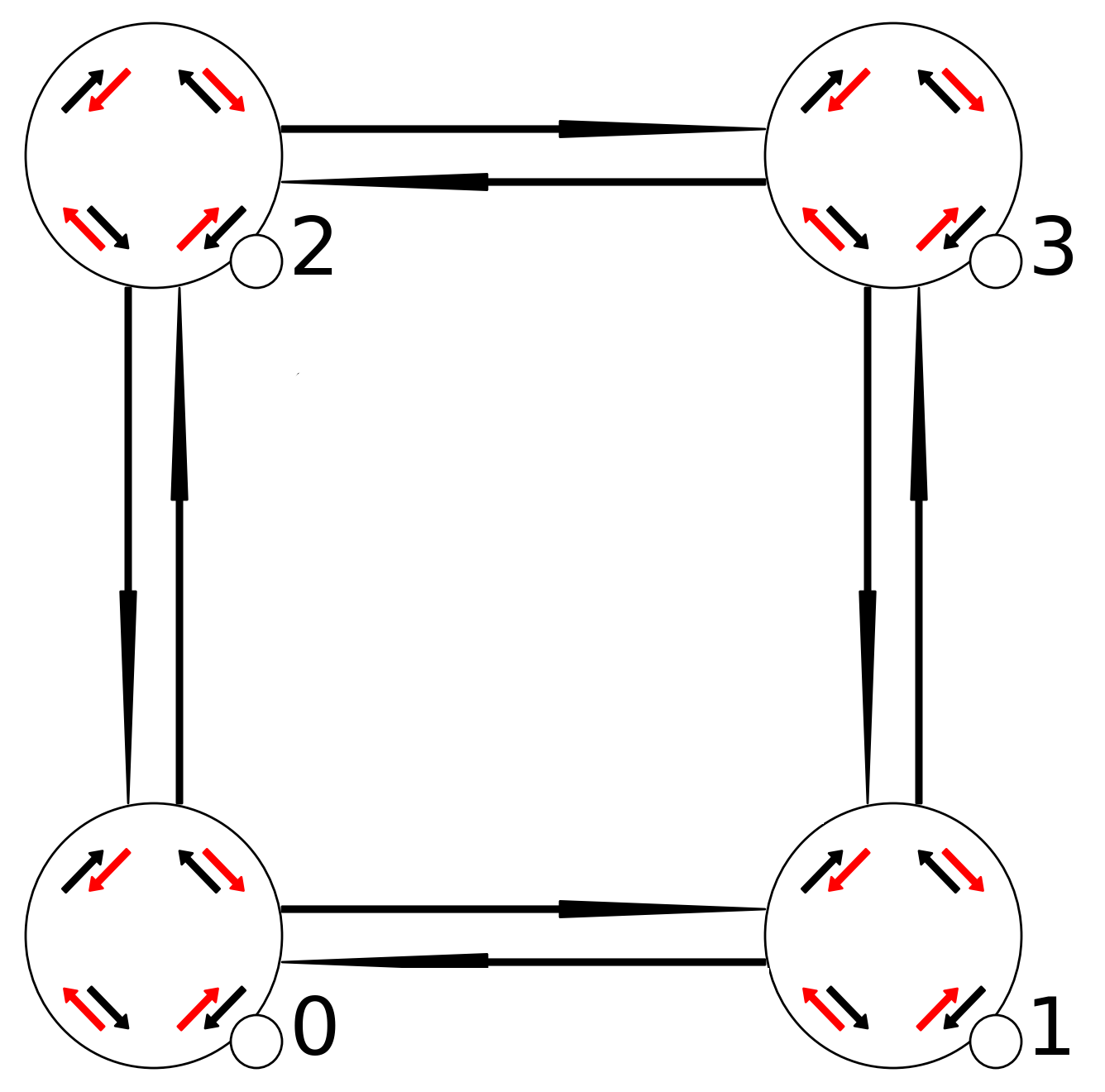}
  \caption{}
  \label{fig:SHM}
\end{subfigure}
\begin{subfigure}{.49\columnwidth}
  \centering
  \includegraphics[width=0.95\columnwidth]{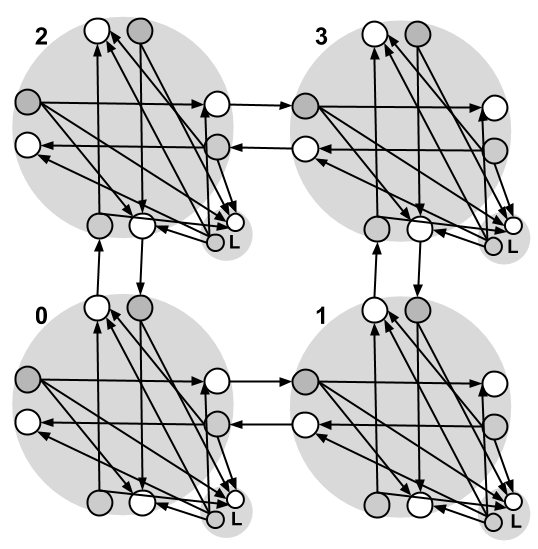}
  \caption{}
  \label{fig:RG}
\end{subfigure}%
\caption{a) Visualization of System Health Map of a $2\times2$ 2D Mesh network with XY routing, b)  Example of Routing Graph for the same network.}
\label{fig:SHMandRG}
\end{figure}

The Mapper-Schedule Unit (MSU) is in charge of mapping and scheduling the application onto the target architecture. A more detailed diagram of MSU is shown in Fig. \ref{fig:MSU}. This unit maintains the following information: \\
\textbf{Application Model}: applications are represented in  form of Task Graphs. A Task Graph is denoted as an acyclic directed graph $TG(V, E)$, in which V represents a set of all tasks $T_i, i \in [0,m]$ (where $m$ shows the number of tasks in application), and E denotes the set of all edges between the tasks which describes the interdependence of the tasks. There is a weight associated with every graph edge that
shows the communication weight between two tasks. Later on, in case of having a big search space for mapping algorithms, if m (the number of tasks) is large, TG is clustered by the MSU into a Clustered Task Graph, CTG(V,E), which is a directed graph where V is a set of clusters in which each cluster contains a set of tasks, and E is the set of edges between two clusters, which encapsulates the communications between tasks in those clusters, using different local-search heuristics. \\
\textbf{Architecture Model}: The NoC topology is described in form of a directed graph, called Architecture Graph (AG) in which each node represents a tile, consisting of a router and Processing Element (PE), and each edge denotes the communication link between two tiles.\\
 \textbf{Routing Model}: Routing is described in form of a Routing Graph (RG) which is an acyclic directed graph, modeling internal and external connections of the routers. The concept of Routing Graph is adopted from \cite{kogge2015}. Fig. \ref{fig:SHMandRG}b shows visualization of an example of such routing graph for a 2x2 Mesh network. Each router consists of five input nodes (four from adjacent routers and one from local) and five output nodes (four to adjacent routers and one towards local). By adding an edge inside the router, a turn or direct connection can be defined for routing algorithm. Adding an edge between two adjacent routers shows a physical link connection between the two. For 3D networks RG becomes slightly more complicated, because instead of ten nodes per router, it requires fourteen nodes per router (two additional nodes for up and two for down direction are required). RG by being acyclic guarantees the absence of deadlocks in the routing algorithm. Using Routing Graph concept enables us to use standard graph algorithms to find routes in the network under any turn model based routing algorithm. A path from a source to destination starts at local input port of the source node and ends at the local output port of the destination node. In case of deterministic routing algorithms such as XY routing, only one path exists in RG from each source to each destination and the packets would be scheduled on the physical links along this route. However, for adaptive routing, it is possible that different routes are returned by RG from each source to destination, in this case scheduling of the packets would be performed probabilistically. The Routing Graph is in sync with SHM, i.e. if a connection fails in the system and is updated in SHM as broken, the same connection is removed from RG. 
 
\subsection{Mapper Memories}
Most Probable Mapping (MPM) memory is used by Mapper-Scheduler Unit (MSU) to keep some mapping configurations for future use to improve the system's response to faults (please refer to section \ref{SHMU} for more details). 
Current Mapping Memory (CMM) is a shared memory between the MSU and SHMU in order to pass the current mapping/scheduling configuration from MSU to SHMU. Information in CMM would be used for determining the impact of the faults on system performance and choosing the proper action after fault occurrence.

In the following sections, the above-mentioned components are described in detail.

\section{Target Architecture}
\label{ArchUnderUse}

The assumed target architecture is a many-core system in which the Processing Elements (PE) communicate using a Network-on-Chip (NoC). This NoC can be a 2D or 3D Mesh or a subset of these topologies (This occurs due to defects in the chip or permanent faults in the links between the routers). 
We describe the fault-prone architecture with an Architecture Graph $Arch: AG(PE(F), C(F))$ where $PE(F)$ is a set of $\{pe_i(F),  \forall i \in [0, N]\}$ 
and $pe_i(F)$ is describing the $ith$ Processing Element under fault configuration F. 
And $C(F)$ is the set of $\{c_j(F), \forall j \in [0, K]\}$  where each $c_j(F)$ is representing the $jth$  communication link under fault configuration F.

In this work we assume that the target architecture supports the following features:

\subsection{Routing re-configuration}
One of the proposals which addresses the implementation and re-configuration of routing algorithms in Network-on-Chips is Logic-Based Distributed Routing (LBDR) \cite{LBDR}. LBDR is a mechanism which is able to implement routing algorithms in 2D NoCs using only logical circuits, thus removing the need for routing tables at each router. This, in turn, leads to better scalability of the mechanism compared to table-based routing approaches. LBDR performs the routing decision based on two sets of configuration bits, known as connectivity and routing bits. The former describes the network topology in form of four bits per router, each one showing connectivity to one of the cardinal directions (North, East, Weat or South), whereas the latter describes the routing algorithm in form of allowed and restricted turns. Routing restrictions are considered as 90-degree turns, thus, 8 bits are considered at each router for the routing bits. It is worth noting that the values of the configuration bits are set offline before the normal operation of the network, therefore, not imposing any additional latency overhead during run-time. This is performed using the framework explained in \cite{OSR-LITE}. 

One main advantage of LBDR over routing tables is that as the size of the network grows, the complexity of the routing logic at each router remains the same, thus being a scalable mechanism. In addition, the values of connectivity and routing bits can be used in order to update information in System Health Map (SHM) memory in SHMU. At each router, if an output port (connected to other routers) is faulty or not active, the corresponding connectivity bit is set to zero. Moreover, if there is a turn restriction at each router or the corresponding turn is faulty in a router, this can be set by modifying the values of the routing bits. It is worth noting that the values of the configuration bits in LBDR only need to be updated once the network has started operating for the first time and also whenever a fault occurs on an output port or in a turn in the router. As mentioned earlier, the (re-)configuration process and (re-)computation of the new values for the bits are performed offline and fed into the logic.

\subsection{Online fault detection mechanism} 

In order to be able to capture the occurrence of faults during run-time in routers, each router is equipped with an online fault detection circuity (known as concurrent online checkers \cite{NOCALERT,CHECKERS}) embedded in it, which basically performs verification of different properties that should hold for different components of the router. It is worth noting that the checkers proposed and used in this work are able to detect both data-path and control part faults in the NoC routers. To be more specific, single Stuck-at-0 (SA0) and Stuck-at-1 (SA1) faults are handled by the checkers.

Regarding the control part, checkers exist which can cover the faults occurred: (1) in the routing computation unit (in our case, the routing logic is implemented using LBDR mechanism), (2) in the arbitration unit and (3) in the control part of input buffers (which are implemented as First-In-First-Out (FIFO)). For the data-path, a single odd parity bit is utilized, providing the detection of single stuck-at faults.

To propagate the fault information to SHMU, iJTAG \cite{IJTAG} mechanism is used. This would enable online and modular access to checkers.
Some tiles of the network are equipped with gateways for passing diagnosis information to the System Health Monitoring Unit. This mechanism enables SHMU to run some test programs on each PE and collect the test data through the NoC infrastructure through these gateways. These gateways are implemented as memory mapped devices sharing a bus with PE of the tile and do not impose much area overhead on the system.

\subsection{NoCDepend} 
In this work we assume that the target architecture supports NoCDepend mechanism proposed in \cite{NOCDEPEND1} for reach-ability calculation. This method stores information about non-reachable parts of the network into a limited number of registers in each router output, represented in form of a rectangular shape and are defined using two corners of the rectangle.  These areas are calculated using Routing Graph (RG) by finding a path between every source and destination and merged and optimized further more. The number of dedicated registers is decided by the user and can vary based on the system's area constraints. Before assigning an output to a packet, the router checks whether the destination is reachable via that output or not. In case the destination is not reachable from any of router's ports, the router should immediately drop the packet in order to prevent injection of non-reachable packets into the system which can cause network congestion. 
The calculated non-reachable areas would be later on transferred to the registers in the routers using iJTAG \cite{IJTAG}. This calculation is indeed computationally expensive, but this process is only needed every time a permanent fault occurs.

The same mechanism can be used for network partitioning  \cite{NOCDEPEND2} for complete traffic separation of Critical from Non-Critical regions in the chip. This along with reconfigurable routing enables using deterministic routing algorithms for critical regions and adaptive routing algorithm for non-critical regions, which improves the network latency compared to the case of using the same routing algorithm for the whole network. 

\section{System Health Monitoring Unit (SHMU)}

\label{SHMU}
In order to make the management of faults possible, the whole system is facilitated with a System Health Monitoring Unit (SHMU), which collects the fault diagnosis reports from the checkers from all routers and based on this information, it maintains a general view of the health status of different network components. 
Based on the information collected from the checkers and online testing, SHMU is able to update the health status information of the network nodes, turns and links in SHM. The status of a network component can be either set to Healthy or Broken in SHM. Also, SHMU records aging of the processing elements and updates the corresponding information in SHM as the components go through aging. 

One of the main roles of SHMU is to classify faults into Transient, Intermittent and Permanent classes.
For intermittent faults, SHMU should predict a time interval until the fault becomes permanent. SHMU should assign a severity metric to the received fault diagnosis information and take appropriate action in order to mitigate the effects. This means that SHMU can either ignore the fault (in case of no severe effect on the system), or can issue an order to Mapper-Scheduler Unit (MSU) to prepare another mapping. This mapping either should be used immediately or in case of prediction of occurrence in short future should be stored in Most Probable Mapping memory.

We define Most Probable Faults Set (MPFS) as a set of $f_i$ most probable faults that can occur to each $AG(PE (F), C(F))$ based on diagnosis information received from the architecture under a certain
mapping $M (App, Arch(F))$. These most probable faults would be stored in the SHMU database. We can define a set of next most probable mappings
as:

 \begin{center}
$ MPM: \{ M(App, Arch(F + f_i)) | \forall f_i \in  MPFS\}$
\end{center}

To calculate these mappings, SHMU updates the System Health Map with a certain fault (which is predicted to happen), then issues a map and store order to the Mapping-Scheduling unit. Upon receiving this, the MSU will calculate the mapping and store it with a fault tag (which encapsulates the fault configuration) in MPM memory. Afterwards, SHMU would return the SHM to its original state.
In terms of the level of granularity, the proposed SHMU in this work is able to deal with faults occurred on Processing Elements (PEs) (node faults), routing turns (turn faults) and communication links (link faults). 

The placement of SHMU in the system is also important, as it can be either one of the nodes in the network, or it can be a separate unit, embedded in the system, but not as part of the NoC. It should be noted that we have opted for the latter implementation since it makes it easier to for SHMU and MSU to communicate through a dual port shared memory (System Health Map). 




\section{Mapper-Scheduler Unit (MSU)}
\label{MSU}
The Mapper-Scheduler Unit (see Fig. \ref{fig:MSU}) is in charge of finding mapping/scheduling solutions for each fault configuration. To perform this task, MSU needs to keep Task Graph (TG) which is essentially the application that encapsulates information about release time, Worst Case Execution Time (WCET) and criticality level of the tasks, and their dependencies, Architecture Graph (AG) which keeps information regarding network topology and portioning, Clustered Task Graph (CTG) which would be used to reduce search space of the mapping problem for large applications and the Routing Graph (RG) which describes how packets can flow through the network. Based on these models, MSU runs mapping and scheduling heuristics (either greedy local search or iterative local search with different cost functions to optimize either the scheduling length, reduce/balance the network traffic, balance processor utilization, etc.-the choice of algorithm and cost function is made by user) and maps TG (or CTG in case of clustering) onto AG. 
The main target of these heuristics is finding a feasible solution with acceptable quality, in few steps. The quality of these heuristics are benchmarked against simulated annealing. However, these heuristics are out of the scope of this paper and would be addressed separately.

The mapping generated by MSU should be in compliance with constraints of RG. We can formulate this process as:

\begin{center}
$M (App, Arch) : TG(V, E) \rightarrow AG(P(F), C(F))$
\end{center}

It does not matter to directly map TG on AG or perform mapping on CTG, in the end a PE would be assigned for each task on TG for execution, that is the main reason that we do not change our representation for mapping for different cases.

Upon receiving a "Map and Store" order from SHMU, MSU starts calculating the mapping by choosing proper heuristic according to information recieved from SHMU. For each mapping, we define a list which $ith$ element of it represents the processing element id chosen for mapping for the $ith$ task. Constructing such a list will provide us with an estimate of the memory usage in the mapper/scheduler. Each mapping also has a fault tag that describes the fault configuration associated with it. This will result in the following data-structure for each mapping in MPM:

\begin{center}
\begin{tabular}{|c|c|c|c|c|}
    \hline
    Fault Tag & $p_{t_0}$ &$p_{t_1}$&...&$p_{t_m}$ \\
 \hline
\end{tabular}
\end{center}
Different approaches can be taken for Fault Tag generation. The easiest is to hash the SHM into a fixed size string. The size of this memory depends on the application size and system requirements. Bigger number of mappings to be stored, require more number of entries in MPM which results in bigger memory size.

Upon receiving a "Map and Deploy" order from SHMU, MSU first checks the fault configuration with the pre-calculated mappings tags in MPM. In case of having a hit in MPM, MSU goes further and extracts the difference with the current mapping in order to perform a partial mapping and then deploys the mapping on PEs. In case of not finding the fault tag in MPM, MSU starts calculating the mapping for the current fault configuration, performs partial mapping extraction and deploys it on PEs.  

We define reconfiguration latency $T_{RL}$ as the time needed for mapper/scheduler to reconfigure the architecture and map the application after a
remap order is received. In cases that the occurred fault is not in MPFS, the reconfiguration latency would be: 

\begin{center}
    $T_{RL} = T_{MapAlg}+T_{ParExt}+T_{ParMap}$
\end{center}

Where $T_{MapAlg}$ is the time needed for computation of a new mapping and scheduling algorithm. $T_{ParExt}$ is the time needed for extracting partial mapping and $T_{ParMap}$ describes the time required for applying partial mapping on the architecture. 
If there is a hit in MPM, the reconfiguration latency $T_{RL}$ would be:

\begin{center}
    $T_{RL} = T_{fetch} + T_{Schd} +T_{ParExt}+T_{ParMap}$
\end{center}

Where $T_{fetch}$ refers to time required to fetch the mapping from the memory and $T_{Schd}$ is needed for calculating scheduling times. $T_{Schd}$  is needed because we are not storing scheduling times for the tasks to save memory. By using  ASAP scheduling algorithm the system would regenerate the same scheduling from the mapping which is of the complexity of $O(n)$.
This would enable the system to recover from a failure  $T_{MapAlg}-(T_{fetch}+T_{Schd})$ time units faster. It is important to note that in case of performing mapping, the system has to run the scheduling algorithm once in each mapping step in order to calculate the cost of mapping since scheduling length is an important factor in the mapping cost function. This would make the gap between these two terms rather significant. Even though these mapping heuristics are fast, they need tens (sometimes hundreds) of steps to find a suitable solution. Another important note is that this improvement is for the best case scenarios, In the cases of MPM miss, the same long mapping scheduling calculation process is needed. Therefore, the performance of this approach is heavily dependant on how well the system can predict the Most Probable Faults. This part is at the moment under development.

\section{Experimentation Setup}
In order analyze the dependability mechanisms in a NoC based many core system and perform the experiments, we have developed an open-source tool, written in Python, named "Schedule and Depend" \cite{GithubRepo} in which the whole target architecture and the proposed components for providing fault-tolerance are implemented in software, along with the possibility of simulating different application scenarios. The tool supports different 2D and 3D topologies and network partitioning based on NoCDepend method, along with the support for different heuristic-based methods (with different cost functions) for clustering and mapping/scheduling of application tasks onto the NoC-based system. In addition, different deterministic and adaptive, turn model based routing algorithms are supported by the tool. The tool provides the user with different possibilities for choosing the application running on the target architecture (from randomly generated Task Graphs to user defined application and some available benchmarks) and has support for mixed-critical applications in which some tasks have higher criticality and must be scheduled within a specified time interval. Finally, the tool is equipped with a Graphical User Interface (GUI) which makes it possible for the user to easily set different parameters, along with visualization of different results obtained from clustering tasks, scheduling, mapping and process of optimization for mapping and clustering heuristics. The project is maintained under GNU-GPL2 license in github, along with all the documentation, available in project wiki. 

\section{Conclusion and Future Work}
\label{Conclusion}
In this paper we proposed a fault-tolerant approach for Network-on-Chip based many-more systems which incorporates a System Health Monitoring Unit that keeps a holistic view of the system health by collecting fault information via online  fault detection mechanisms using iJTAG chains. The fault information is classified into different classes (transient, intermittent and permanent), the Mapping/Scheduling Unit is utilized in order to find feasible mapping scheduling solutions for system recovery. A mapping storage system based on prediction of future faults is proposed to reduce the reconfiguration time after each system failure. Currently, all the proposed components are developed in our system except fault classification/prediction which is currently under development. By completion of this unit, the system loop would be closed and some comparisons could be made with the state of the art. An experimentation environment is developed for simulation of the proposed approach and a Field Programmable Gate Array (FPGA) based emulator is also under development in order to test the proposed methods and approaches on hardware. This would confirm if the scheduling and simulation results would comply with the actual implementation of the system when different applications are running on the system.


%


\section*{Acknowledgment}

The work has been supported by EU FP7 STREP BASTION, EU's H2020 RIA IMMORTAL, Estonian Science Foundation grant ETF9429, Estonian institutional research grant IUT 19-1, and funded by Estonian Ministry of Education and Research.

\bibliographystyle{IEEEtran}
\bibliography{IEEEabrv,biblio.bib}

\end{document}